\documentclass[aps,prc,superscriptaddress,showpacs,nofootinbib,floatfix,twocolumn]{revtex4}

\usepackage{graphicx}   
\usepackage{rotating}
\usepackage{dcolumn}
\usepackage{bm}
\usepackage{epsfig}
\usepackage{amssymb}
\usepackage{multirow}

\begin{document}
\title{Influence of the nucleon-nucleon collision geometry on the determination of the nuclear modification factor for nucleon-nucleus and nucleus-nucleus collisions}
\newcommand{\sunysb}{Department of Chemistry, Stony Brook University, Stony Brook, NY 11794, USA}
\newcommand{\bnl}{Physics Department, Brookhaven National Laboratory, Upton, NY 11796, USA}
\author{Jiangyong Jia}
\affiliation{\sunysb}\affiliation{\bnl}\date{\today}

\begin{abstract}
The influence of the underlying nucleon-nucleon collision geometry
on evaluations of the nuclear overlap function ($T_{AB}$) and
number of binary collisions ($N_{coll}$) is studied. A narrowing of
the spatial distribution of the hard-partons with large light-cone
fraction $x$ in nucleons leads to a downward correction for
$N_{coll}$ and $T_{AB}$, which in turn, results in an upward
correction for the nuclear modification factor $R_{AB}$. The size
of this correction is estimated for several experimentally
motivated nucleon-nucleon overlap functions for hard-partons. It is
found to be significant in peripheral nucleus-nucleus and
nucleon-nucleus collisions, and are much larger at the LHC energy
of $\sqrt{s}=5.5$ TeV than for the RHIC energy of $\sqrt{s}$=200
GeV. The implications for experimental measurements are also
discussed.
\end{abstract}

\maketitle

\section{Introduction}
In experiments at the Relativistic Heavy Ion Collider (RHIC),
modification of hard-scattering processes is usually quantified via
the nuclear modification factor. For collision between nucleus A
and B (A-B collision), it is defined as
\begin{eqnarray}
\label{eq:1}
R_{AB}=\frac{\frac{1}{N_{\rm evt}^{\rm AB}}\frac{dN_{\rm AB}}{dp_T}}{\langle T_{AB}\rangle \frac{d\sigma_{nn}}{dp_T}}
= \frac{\frac{1}{N_{\rm evt}^{\rm AB}}\frac{dN_{\rm AB}}{dp_T}}{\langle N_{\rm coll}\rangle \frac{1}{N_{\rm evt}^{\rm nn}}\frac{dN_{\rm nn}}{dp_T}}
\end{eqnarray}
where $T_{AB}$ is the nuclear overlap function calculated as the
convolution of the thickness functions $T_{\rm A,B}({ \vec b}) =
\int \rho_{\rm A,B}({ \vec b},z)dz$ for A and B,
\begin{eqnarray}
\label{eq:2}
T_{\rm AB} ({ \vec b})=\int d{ \vec s}\;
T_{A}({ \vec s})T_{B}({ \vec s}-{ \vec b}),
\end{eqnarray}
and ${\langle T_\mathrm{\rm AB}\rangle}$ is the average nuclear
overlap function for the corresponding centrality bin
\begin{equation}
\label{eq:3}
{\langle T_{\rm AB}\rangle}\equiv
\frac {\int T_{\rm AB}({ \vec b})(1- e^{-
\sigma^{\rm inel}_{nn}\, T_{\rm AB}({ \vec b})}) \, d{ \vec b} }{\int (1- e^{-
\sigma^{\rm inel}_{nn}\, T_{\rm AB}({ \vec b})})\, d{ \vec b}}=
\langle N_{\rm coll}\rangle/\sigma^{\rm inel}_{nn} \quad.
\label{eq:RAA}
\end{equation}
In Eqs.~\ref{eq:1} and \ref{eq:3}, $\langle N_{\rm coll}\rangle$ is
the average number of inelastic nucleon-nucleon (n-n) collisions
with inelastic n-n cross section $\sigma^{\rm inel}_{nn}$. At the
nominal RHIC energy of $\sqrt{s}=200$ GeV,
$\sigma_{nn}^{inel}=42mb=4.2fm^2$. The estimation of $N_{coll}$,
$T_{AB}$ and other geometrical quantities are usually obtained
either numerically via an optical Glauber approach or statistically
via a Monte-Carlo Glauber approach (see Appendix~\ref{app:A}). More
details on the Glauber model can be found in~\cite{Miller:2007ri}.

\begin{figure*}[ht]
\epsfig{file=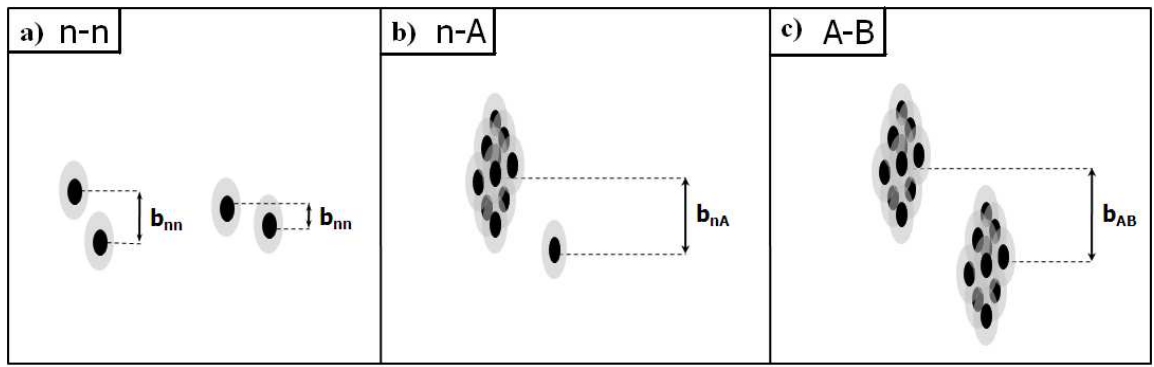,width=0.85\linewidth} \caption{\label{fig:1}
Schematic illustration of the collision geometry for
nucleon-nucleon or n-n (a), nucleon-nucleus or n-A (b) and
nucleus-nucleus or A-B (c) collisions. a) Collision geometry for
peripheral (left) and central (right) n-n collisions. The dark
(light) region indicates the transverse distribution of
hard-partons and soft-partons, respectively (figure adapted
from~\cite{Frankfurt:2003td}), hence the hard-scattering rate is
suppressed in peripheral collisions. b) Collision geometry for
peripheral n-A collisions. The nucleus only overlap with top half
of the projectile nucleon, hence the corresponding n-n collisions
is bias towards large n-n impact parameter. c) Collision geometry
for peripheral A-B collisions. The corresponding n-n impact
parameter distribution receive surface biases from both nucleuses.}
\end{figure*}
The definition of $T_{AB}$ in Eq.~\ref{eq:2} ignores the fact that
nucleons are extended object and the multiplicity of a n-n
collision also depends on the impact parameter $b_{\rm nn}$. This
can be accounted for by introducing a n-n overlap function $t({
\vec b_{nn}})$ with normalization of $\int d{ \vec b_{nn}} t({ \vec
b_{nn}})=\int {db_{nn}2\pi b_{nn}\;} t(b_{nn})=1$, which
generalizes Eq.~\ref{eq:2} to Wong's formula~\cite{Wong:1984sy}:
\small{
\begin{eqnarray}
\label{eq:4}
\nonumber
T_{\rm AB} ({ \vec b_{AB}})&=&\int d{ \vec b_A}d{ \vec b_B}\;
T_{A}({ \vec b_A})T_{B}({ \vec b_B})t({ \vec b_{AB}}- { \vec b_A}+{ \vec b_B})\\\nonumber
&=&\int d{ \vec s}d{ \vec b_{nn}}\;
T_{A}({ \vec s})T_{B}({ \vec s}-{ \vec b_{AB}}+ { \vec b_{nn}})t({ \vec b_{nn}}).\\
\end{eqnarray}
}\normalsize In this notation, Eq.~\ref{eq:2} is obtained for the
special case when the n-n overlap function is a delta function:
$t(b_{nn})=\delta(b_{nn})$. Eq.~\ref{eq:4} shows that the nuclear
overlap function $T_{\rm AB}$ (and thus $N_{\rm coll}$) depends on
not only the nuclear impact parameter $b_{\rm AB}$, but also the
n-n impact parameter $b_{nn}$ through $t(\vec b_{nn})$. In other
words, the $b_{\rm AB}$ and $b_{\rm nn}$ are correlated. Because of
this correlation, Eq.~\ref{eq:4} in general is not the same as
Eq.~\ref{eq:2} at fixed $b_{AB}$. However if the nuclear thickness
function varies linearly within the length scale of nucleon size,
Eq.~\ref{eq:2} is a good approximation since the integral of the
first term of its Taylor expansion vanishes due to spherical
symmetry of the nucleon,
\begin{eqnarray}
\label{eq:5}
\nonumber
&&\int d{ \vec b_{nn}}\;T_{B}({
\vec s}-{ \vec b_{AB}}+{ \vec b_{nn}})t({ \vec b_{nn}})\\\nonumber
&&\quad\quad\approx
\int d{ \vec b_{nn}}\left[T_{B}+\nabla T_{B}\cdot { \vec
b_{nn}}\right]t({ \vec b_{nn}})=T_{B}({ \vec s}-{ \vec b_{AB}}).
\end{eqnarray}
This approximation is rather precise for central A-B collisions,
but breaks down in peripheral collisions. Furthermore one can show
that Eq.~\ref{eq:4} (hence Eq.~\ref{eq:2}) obeys the following sum
rule
\begin{eqnarray}
\label{eq:6}
\nonumber
&&\int d{ \vec b_{AB}} T_{AB}({ \vec b_{AB}})\\\nonumber
&&=
\int d{ \vec s}d{ \vec b_{nn}}d{ \vec b_{AB}}\; T_{A}({ \vec s})T_{B}({ \vec
s}-{ \vec b_{AB}}+{ \vec b_{nn}})t({ \vec b_{nn}})\\\nonumber
&&=\int d{ \vec s}d{ \vec b_{AB}}\;
T_{A}({ \vec s})T_{B}({ \vec s}-{ \vec b_{AB}})\\
&&=AB/\sigma_{AB}^{geo},
\end{eqnarray}
where $\sigma_{\rm AB}^{\rm geo}$ is the total A-B geometrical
cross-section. This equation implies that the integral of $T_{\rm
AB}$, i.e. for minimum bias events, is independent of the n-n
overlap function.

The transverse spatial distribution of the partons in a highly
lorentz-boosted nucleon depends on the light-cone fraction
$x$~\cite{Frankfurt:2003td,Frankfurt:2005mc}. The distribution of
hard-partons ($x\gtrsim10^{-2}$) in the transverse plane (xy-plane)
is considerably narrower than that for soft-partons which define
the geometrical size of the nucleon (see
Fig.1a)~\cite{Weiss:2009ar}. The reason is that the transverse
profile of soft-partons grows with $ln(1/x)$ due to Gribov
diffusion~\cite{Gribov:1973jg}, corroborated by the possible
tamping of the growth of soft-partons at nucleon center by
saturation effects~\cite{Iancu:2003xm}. As a consequence, the n-n
impact parameter distribution, i.e. the overlap function
$t(b_{nn})$, is narrower for hard-scattering processes than that
for minimum-bias n-n collisions. If we denote the n-n overlap
function for hard-scattering processes as $t_{hs}(b_{nn})$, the
corresponding nuclear overlap function and equivalent number of n-n
collisions for Eq.~\ref{eq:1} has a form similar to Eq.~\ref{eq:4}
(see Appendix~\ref{app:B}):
\begin{eqnarray}
\label{eq:7}
\nonumber
T_{\rm AB}^{hs} ({ \vec b_{AB}})&=&\int d{ \vec s}d{ \vec b_{nn}}
T_{A}({ \vec s})T_{B}({ \vec s}-{ \vec b_{AB}}+ { \vec b_{nn}})t_{hs}({ \vec b_{nn}})\\
N_{coll}^{hs}  &=& T_{AB}^{hs} \sigma_{nn}^{inel}
\end{eqnarray}
The primary goal of this paper is to investigate the sensitivity of
$T_{AB}$ and $N_{coll}$ to different assumptions about n-n overlap
function for hard-scattering processes. The influence of this
sensitivity on the interpretation of the nuclear modification
factor $R_{AB}$ is also discussed.

\section{Results and Discussion}
To help visualizing the level of correlation between $b_{AB}$ and
$b_{nn}$, we plot the $b_{nn}$ distributions for various ranges of
$b_{AB}$ in Au-Au collisions in Figure~\ref{fig:2}. These
distributions are calculated using the MC Glauber approach (see
Appendix~\ref{app:A}). A n-n total inelastic cross section of
$\sigma_{nn}^{inel}=42$ mb is used in the calculation, which
corresponds to a geometrical radius of
$r_{n}=\sqrt{\sigma_{nn}^{inel}/\pi}/2=0.5781fm$ for nucleons. A
n-n collision is considered to occur when their distance in the
xy-plane is less than $d_{max}=2 r_n=1.156fm$. The individual
distributions in Figure~\ref{fig:2} have been re-scaled to match
each other at $b_{nn}$=2.5fm. For minimum bias Au-Au events, the
distribution increase linearly with $b_{nn}$. This is so because
all possible xy positions of a given nucleon in one nucleus
relative to the second nucleus are uniformly sampled (consistent
with the sum rule Eq.~\ref{eq:6}). Thus the corresponding $b_{nn}$
distribution has the same shape as that for pure n-n collisions. By
contrast, the distributions for centrality selected Au-Au events
are not linear functions of $b_{nn}$. For peripheral Au-Au
collisions, the n-n distribution has a concave-like shape, implying
that the selection of large $b_{AB}$ results in a bias towards
peripheral n-n collisions (large $b_{nn}$). For central Au-Au
collisions, the n-n distribution has a slight convex-like shape,
implying that the requirement of a small $b_{AB}$ value leads to a
bias towards slightly central n-n collisions.
\begin{figure}[h]
\epsfig{file=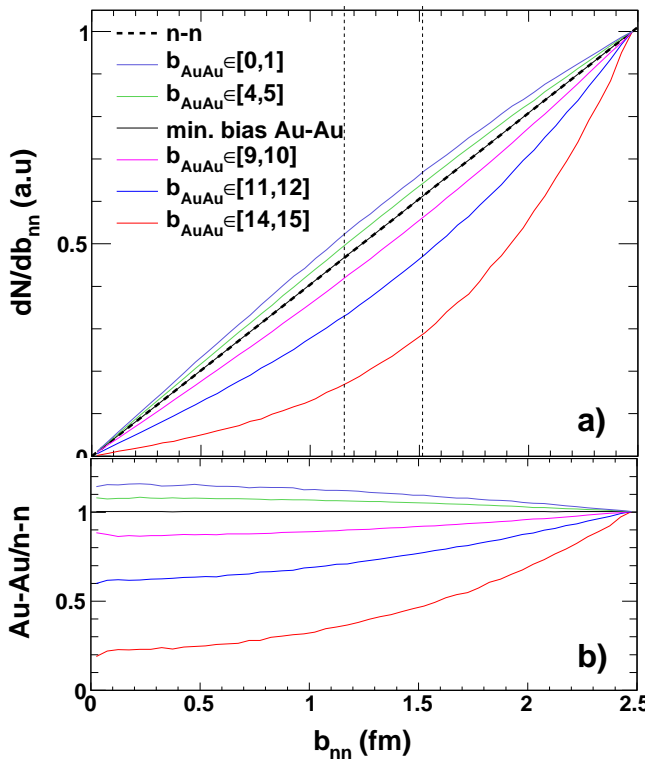,width=1\linewidth} \caption{\label{fig:2}
a) Nucleon impact parameter distributions of n-n collisions for
centrality selected Au-Au events. From top to bottom, the curves
show several Au-Au impact parameter selections, ranging from 0-1 fm
to 14-15 fm. The minimum bias Au-Au distribution and pure n-n
distribution fall on top of each other. The two vertical dashed
lines at 1.156fm and 1.514fm indicate the radius of the n-n
inelastic cross-section at $\sqrt{s}$=200 GeV (42mb) and 5.5 TeV
(72mb), respectively. b) The ratios between various centrality
selected Au-Au $b_{nn}$ distributions to n-n.}
\end{figure}

The origin of the bias can be further illustrated via peripheral
n-A collisions. According to Eq.~\ref{eq:4}, the n-n impact
parameter distribution for fixed $b_{nA}$ should be
\begin{eqnarray}
\label{eq:den}
\nonumber
f(b_{nn}, b_{nA})&=& \int d{ \vec{s}}d\phi\; T_{B}({ \vec{s}})T_{A}({\vec{s}-\vec{b}_{nA}+\vec b_{nn}})\; b_{nn}\\
&=&\int d\phi\; T_{A}(\vec{b}_{nA}-\vec b_{nn})\; b_{nn},
\end{eqnarray}
after exchanging subscript A and B and setting $T_{B}=\delta(\vec
s)$. When the distance of nucleon from the nuclear surface is of
the order of nucleon size, i.e. $|b_{nA}-R|\lesssim r_{n}$ (R is
the nuclear radius and $r_n=d_{max}/2$ is nucleon radius), the
projectile nucleon see many more nucleons at the inner side of the
target nucleus than the side close to the surface (see
Figure~\ref{fig:1}b), leading to the bias shown in
Figure~\ref{fig:2}. This bias is larger for A-B collisions, since
the overlap function $T_{AB}$ is the convolution of two thickness
functions, which varies more rapidly towards the edge of the
overlap region. Furthermore, the magnitude of such bias should also
be sensitive to the nuclear profile. Due to the surface diffuseness
of the Woods-saxon nuclear profile (used in this analysis), the
range of $b_{AB}$ affected by this bias is expected to be larger
than that for hard-sphere nuclear profile.

\begin{figure}[h]
\epsfig{file=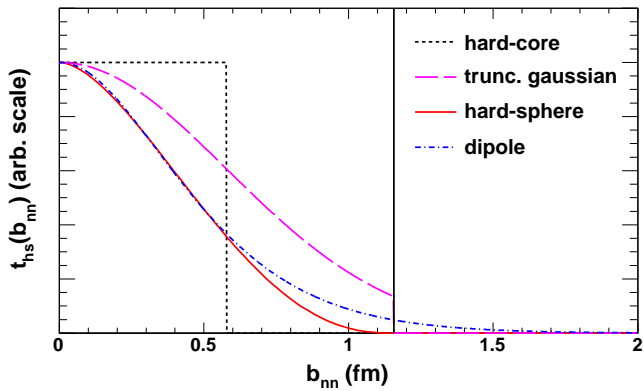,width=1.0\columnwidth}
\caption{\label{fig:3}
The shapes of the impact parameter distribution of the overlap
function for hard-scattering processes at $\sqrt{s}=200$ GeV; the
solid vertical line indicates the range of impact parameters for
the minimum bias n-n condition. For tje LHC energy of
$\sqrt{s}$=5.5 TeV, we simply increase the nucleon radius $r_n$
from 0.578fm to 0.757fm, which stretches all distributions (except
the dipole formula) horizontally by 30\%.}
\end{figure}

Measurements of hard inclusive scattering processes have provided a
detailed picture of the longitudinal momentum distribution of
partons in a nucleon. Unfortunately, information on the transverse
distribution of partons in a nucleon, or the impact parameter
dependence of the hard-scattering process in n-n collisions, is
rather limited. In general, it is a function of $x$ and $Q^2$ and
can be described by projecting the generalized parton distribution
function (GPD)~\cite{Weiss:2009ar,Ji:2003ak}. To estimate the
magnitude of the bias, we tried several different n-n overlap
functions. The default overlap function is a step function defined
as
\begin{eqnarray}
\label{eq:mb}
t_{mb}(b_{nn})\propto\theta(b_{nn}-2r_{n}),
\end{eqnarray}
which treats all n-n collisions on equal footing in calculating the
number of minimum bias n-n collisions. It is the default approach
commonly used in MC Glauber calculations of $T_{AB}$ and
$N_{coll}$~\cite{Miller:2007ri}. We compare it to the following
four parameterizations of hard-scattering overlap functions (see
Figure~\ref{fig:3}), all of which have a width narrower than the
default. The first one assumes that the hard-scatterings are
restricted to a core region which is half the size of the n-n
overlap, i.e.
\begin{eqnarray}
\label{eq:hs1}
t_{hs}^1(b_{nn})\propto\theta(b_{nn}-r_{n}).
\end{eqnarray}
This overlap profile has no rigorous physics foundation, other than
the fact that it provides a simple and intuitive estimation of the
bias. The second one assumes that the overlap function is the
folding of two Gaussian profiles for hard-partons with the width
$r_{n}$.
\begin{eqnarray}
\label{eq:hs2}
t_{hs}^2(b_{nn})\propto \frac{1}{2r_{n}^2} e^{-\frac{b_{nn}^2}{2r_{n}^2}}\theta(b_{nn}-2r_{n}).
\end{eqnarray}
This is the n-n overlap function used by
PYTHIA~\cite{Sjostrand:1987su}, except that we truncate the overlap
function at $2r_{n}$, so the average width is narrower than the
default (about $0.7 r_{n}$ in this case). The third one assumes
that the hard-partons are uniformly distributed in a hard-sphere
nucleon with radius $r_{n}$, the corresponding overlap function is
\begin{eqnarray}
\label{eq:hs3}
\nonumber t_{hs}^3(b_{nn}) \propto\int d\vec s &&\sqrt{1-\frac{s^2}{r_n^2}} \theta(s-r_{n})\times \\\nonumber &&\sqrt{1-\frac{(\vec s-\vec b_{nn})^2}{r_n^2}}\theta(|\vec s-\vec b_{nn}|-r_n).\\
\end{eqnarray}
The last one is the dipole formula for gluons spatial distribution
taken from~\cite{Frankfurt:2003td}. It is derived from fits to
$J/\Psi$ photo-production data at HERA and FNAL,
\begin{eqnarray}
\label{eq:hs4}
t_{hs}^4(b_{nn}) \propto \frac{m_g^2}{12\pi}\left(\frac{m_gb_{nn}}{2}\right)^3K_3(m_gb_{nn}).
\end{eqnarray}
where $K_3$ is the modified Bessel function, and the mass parameter
$m_g^2\sim1.1GeV^2$, which depends weakly on $Q^2$ and assumed to
be constant in this analysis.

\begin{figure*}[h]
\epsfig{file=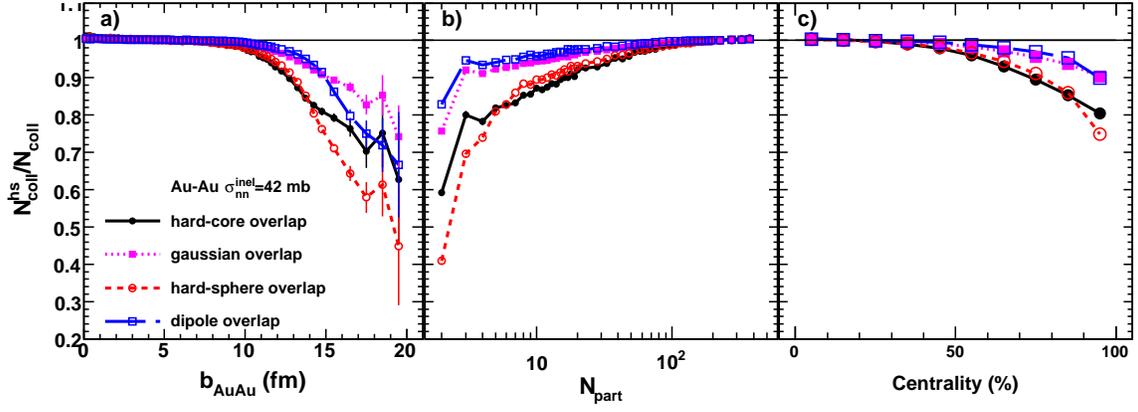,width=0.85\linewidth}
\caption{\label{fig:4} The ratio of $N_{coll}^{hs}$ to $N_{coll}$
calculated for Au-Au collisions for the four n-n overlap functions
described by Eq.~\ref{eq:hs1}-\ref{eq:hs4} plotted as function of
a) impact parameter $b_{AuAu}$, b) $N_{part}$ and c) centrality
bins in 10\% step, assuming n-n inelastic cross section of 42mb
(for RHIC energy).}
\end{figure*}

With these input distributions in hand, it is fairly straight
forward to evaluate the resulting differences in $N_{coll}^{hs}$
and $N_{coll}$ using the MC Glauber approach. The results obtained
for Au-Au collisions are shown in Figure~\ref{fig:4}; they are
calculated for Eq.~\ref{eq:mb} with $\sigma_{nn}^{inel}=42$ mb (for
$\sqrt{s}$=200 GeV). The results are plotted against either a)
Au-Au impact parameter, b) $N_{part}$, or c) centrality in 10\%
steps (sliced according to Au-Au geometrical cross-section). Here,
we shall focus the discussion on the middle panel. One can see that
the $N_{coll}^{hs}$ is always smaller than the nominal $N_{coll}$
value for small $N_{part}$. This is easily understood since
$N_{coll}$ is calculated for a flat probability distribution while
all others are calculated for a distribution which peaks at small
$b_{nn}$. The bias is sizable at $N_{part}<20-50$, corresponding to
$>60$\% centrality range. The bias for the Gaussian profile is
smallest since it's $t_{nn}$ is broader than for the other
scenarios. For the more realistic case of hard-sphere and dipole
overlap, the bias grows to about 5-15\% for $N_{part}\sim 10-15$
which corresponds roughly 70-80\% centrality bin. Note that the
corrections all cross at $N_{part}\sim 150$ (or $b_{AuAu}\sim8$ fm)
and stay slightly above unity for central collisions. This is
expected since there is no bias for minimum bias Au-Au selection,
as mandated by the sum rule Eq.~\ref{eq:6}.

\begin{figure*}[h]
\epsfig{file=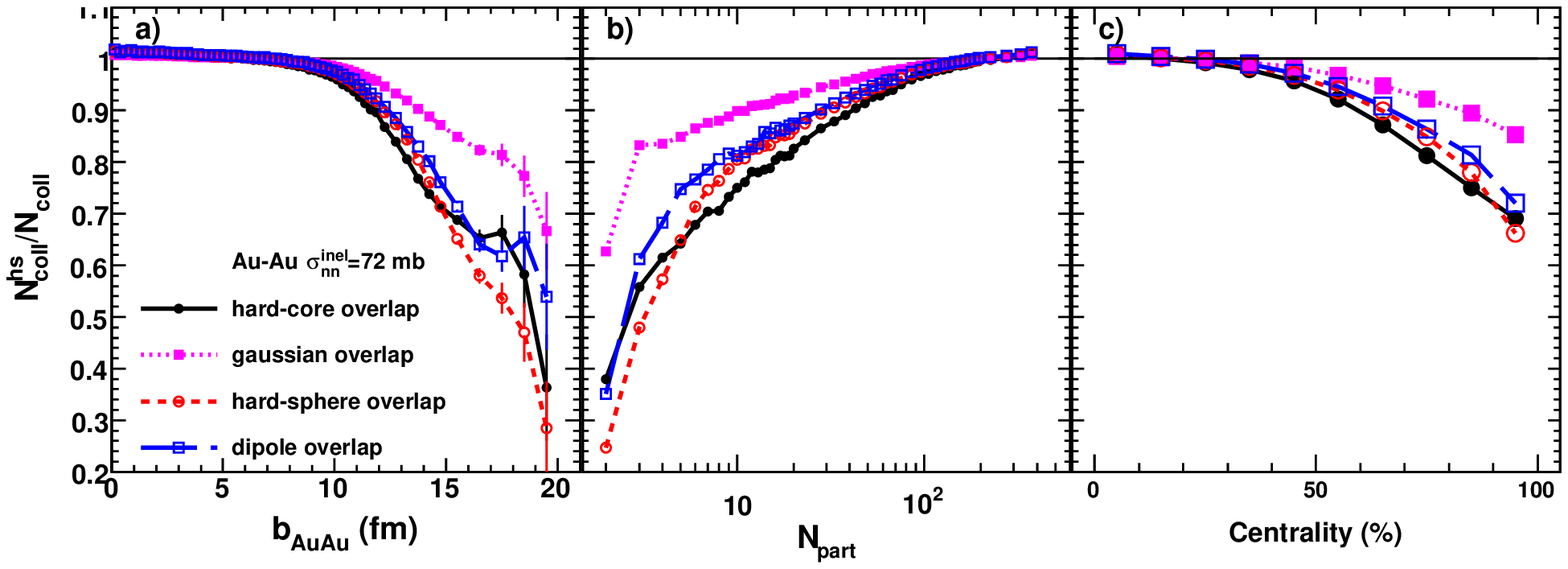,width=0.85\linewidth}
\caption{\label{fig:5} The ratio of $N_{coll}^{hs}$ to $N_{coll}$
calculated for Au-Au collisions for the four n-n overlap functions
described by Eq.~\ref{eq:hs1}-\ref{eq:hs4} plotted as function of
a) impact parameter $b_{AuAu}$, b) $N_{part}$ and c) centrality
bins in 10\% steps, assuming n-n inelastic cross section of 72mb
(for LHC energy).}
\end{figure*}

\begin{figure*}[h]
\epsfig{file=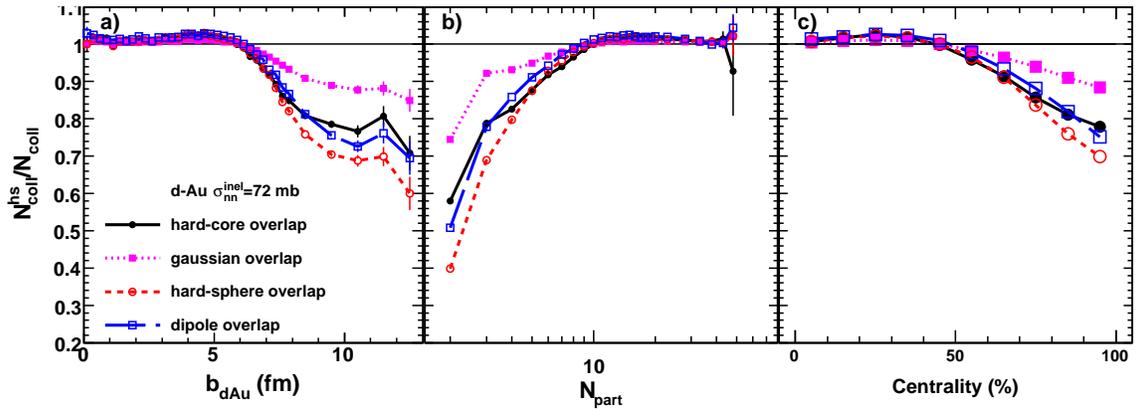,width=0.85\linewidth}
\caption{\label{fig:6} The ratio of $N_{coll}^{hs}$ to $N_{coll}$
calculated for deuteron-gold (d-Au) collisions for the four n-n
overlap functions described by Eq.~\ref{eq:hs1}-\ref{eq:hs4}
plotted as function of a) impact parameter $b_{AuAu}$, b)
$N_{part}$ and c) centrality bins in 10\% steps, assuming n-n
inelastic cross section of 72mb (for LHC energy). The Hul\'then
wave function for deuteron is used. }
\end{figure*}

Based on the trends shown in Figure~\ref{fig:2}, it is clear that
the bias should grow with an increase of total n-n inelastic cross
section. At the LHC energy of 5.5 TeV, the total n-n cross section
is estimated to be $\sigma_{nn}^{inel}\approx72
mb=7.2fm^2$~\cite{d'Enterria:2003qs}, about 75\% higher than that
at the nominal RHIC energy. This corresponds to $d_{max} = 1.514fm$
and $r_{n}=0.757fm$. The results obtained for this condition are
summarized by Figure~\ref{fig:5}, which shows about 60\% larger
bias on $N_{coll}$ when compared to that for the RHIC energy.

As discussed before, the same bias should also exist for p-A
collisions, but to a lesser extent. In Figure~\ref{fig:6}, we show
the results of the calculation for deuteron-Au (d-Au) collisions at
$\sqrt{s}=5.5$ TeV. Indeed the bias is smaller compared to Au-Au
(Figure~\ref{fig:5}) at the same $N_{part}$. However the available
range of $N_{part}$ covered by d-Au collision is small. So the
correction factor for given percentile centrality bin, sliced
according to geometrical cross-section, turns out to be larger. For
the $\langle N_{part}\rangle$ values for 60-100\% centrality range
is only about 4, for which the correction could be as large as
15\%.

The results in Figure~\ref{fig:4}-\ref{fig:6} underlines the
importance of the nucleon-nucleon collision geometry for proper
interpretation of the collision geometry in p-A and A-A collisions.
Depending on the assumed hard-parton profile, the bias on the
$N_{coll}$ (hence $T_{AB}$) could become sizable, especially at LHC
energy where the total n-n cross-section is significantly larger,
and in peripheral p-A and A-A collisions where the correlation
between $b_{AB}$ and $b_{nn}$ is important. In such cases, the full
formula for $T_{AB}$ (Eq.~\ref{eq:4}) and proper n-n overlap
function for hard-partons are needed for proper evaluation of the
nuclear modification factor. The magnitude of the bias is usually
smaller than the typical systematic error quoted by the RHIC
experiments~\cite{Miller:2007ri}. Also the experimental triggering
efficiency for peripheral A-A and p-A collision is low, for
example, the PHENIX experiment triggers on 92\% of Au-Au
collisions~\cite{Adler:2003au} and 88\% of the d-Au
collisions~\cite{Adler:2007by}, hence the relevant correction
factor for experimentally accessible centrality range could be
30-50\% smaller than what has been calculated here. We should point
out that a related bias has been considered before by the PHENIX
Collaboration~\cite{Miller:2007ri}, where a Gaussian profile was
used as part of the estimation of the systematic errors on
$N_{coll}$. However that study was not motivated by the narrowing
of hard-parton spatial distribution, rather it was used to account
for uncertainty of the nucleon matter distribution which are
dominated by soft partons.

In summary, we discussed a bias in the calculation of the
$N_{coll}$ and $T_{AB}$, and its influences on the nuclear
modification factor $R_{AB}$ within the Glauber formulism. The bias
is caused by the difference in the spatial profile between the
hard-partons which contribute to the high $p_T$ yield, and
soft-partons which determines the n-n inelastic cross-section. The
much narrower spatial profile of the hard-partons biases the
$N_{coll}$ and $T_{AB}$ to values smaller than those obtained using
the uniform n-n overlap function. The magnitude of the bias is
sensitive to width of the overlap function for hard-partons. A
crude estimation for Au-Au collisions at RHIC energy leads to about
5\% (assuming a Gaussian overlap function)-15\% (assuming a
hard-sphere overlap function) downward corrections for 70-80\%
centrality bin and becomes significantly larger for more peripheral
collisions, which may account for part of the suppression seen for
very peripheral Au-Au $\pi^0$ data~\cite{Adare:2008qa}. For
$N_{part}>150$, the bias is positive, but is less than a few
percent. Since the bias increases with the total n-n inelastic
cross-section used in the calculation. We estimate the bias could
be 60\% larger at LHC energy compare to RHIC energy, and has to be
properly taken into account in Glauber calculation for both Pb-Pb
and p-Pb collisions.

The author wishes to thank Roy Lacey, Michael Tannenbaum and
Cheuk-Yin Wong for valuable discussions. This research is supported
by NSF under award number PHY-0701487.
\appendix
\section{Optical vs MC approach}
\label{app:A} The calculations of Glauber variables, including
$\langle T_{\rm AB}\rangle$ and $\langle N_{\rm coll}\rangle$, are
carried out either numerically using a set of equations like
Eq.~\ref{eq:2}-\ref{eq:4} (``optical approach'') or statistically
via Monte-Carlo method (``MC approach''). The MC approach is what
has been used by RHIC experimentalist. Comparing with Optical
approach, it takes into account position fluctuations of nucleons
in the nucleus, which is shown to be important in peripheral
collisions. In the MC approach, A-B collisions are generated
randomly in the xy-plane. Within each event, nucleus A and B are
populated randomly with nucleons according to the Woods-Saxon
nuclear profile. All possible n-n combination between the two
nucleus are considered. In the simplest version at RHIC energy, a
n-n collision is considered to happen when their distance in the
xy-plane is less than
$d_{max}=\sqrt{\sigma_{nn}^{inel}/\pi}=1.156fm$ (hard-sphere
assumption, corresponding to $t({ b_{nn}})=\theta(|{
b_{nn}}|-d_{max})$). The number of n-n collisions in a given event
is
\begin{eqnarray}
\label{eq:app1}
\nonumber
N_{\rm coll}  = \sum\limits_{i\in A,j\in B} t(\left| {\vec r_i  -
\vec r_j }\right|)=\sum\limits_{i\in A,j\in B} {\theta\left(\left|
{\vec r_i  - \vec r_j } \right|- d_{\max }\right)}\\
\end{eqnarray}
The $T_{\rm AB}$ is then calculated as $T_{\rm AB}=N_{\rm
coll}/\sigma_{nn}^{inel}$. The $\langle N_{\rm coll} \rangle$ and
$\langle T_{\rm AB} \rangle$ are obtained by averaging over many
events falling in a given centrality definition. More details can
be found in~\cite{Miller:2007ri}.

\section{Scale Factor for Hard-scattering Process} \label{app:B}
In this section, we derive the expression for correct scaling
factor, $N_{coll}^{hs}$, for hard-scattering process. The average
hard-scattering yield in A-B event at a fixed impact parameter can
be expressed as
\begin{eqnarray}
\label{eq:app2}
Y_{AB}^{hs}&=&\langle \sum\limits_{i\in A,j\in B} t_{hs}(|\vec b_{i}-\vec b_{j}|)\sigma_{pp}^{hs}\;\;\rangle_{evts}\\\nonumber
&=&\sigma_{nn}^{hs} \int db_{nn}\; f(b_{nn},b_{AB})t_{hs}(b_{nn})\\\nonumber
&=&\sigma_{nn}^{hs} \int d{ \vec s}d{ \vec b_{nn}}\;
T_{A}({ \vec s})T_{B}({ \vec s}-{ \vec b_{AB}}+ { \vec b_{nn}})t_{hs}({\vec b_{nn}}).\\
&=& \sigma_{nn}^{hs} T_{AB}^{hs}(\vec b_{AB})
\end{eqnarray}
where $f(b_{nn},b_{AB})$ is n-n collision density in impact
parameter space defined by Eq.~\ref{eq:den}, and
\begin{eqnarray}
\label{eq:app3}
T_{AB}^{hs}=\int d{ \vec s}d{ \vec b_{nn}}\;
T_{A}({ \vec s})T_{B}({ \vec s}-{ \vec b_{AB}}+ { \vec b_{nn}})t_{hs}({\vec b_{nn}})
\end{eqnarray}
has the same form as Eq.4 except that the overlap function is
replaced with the one for hard-scattering process.

The n-n hard-scattering rate is simply the hard-scattering profile
averaged over all n-n events in vacuum,
\begin{eqnarray}
\label{eq:app4}
\langle Y_{nn}^{hs}\rangle =\frac{\sigma_{nn}^{hs}\int {db_{nn}2\pi
b_{nn}\;}  t_{hs}(b_{nn})}{\int db_{nn} 2\pi b_{nn}}
           =\frac{\sigma_{nn}^{hs}}{\sigma_{nn}^{inel}}
\end{eqnarray}
where the normalization $\int {db_{nn}2\pi b_{nn}\;}
t_{hs}(b_{nn})=1$ is used.

The correct scaling factor should be the ratio of the
hard-scattering rate in centrality selected A-B collisions to that
of the minimum bias n-n collisions:
\begin{eqnarray}
\label{eq:app5}
N_{coll}^{hs}  &=& \frac{Y_{AB}^{hs}}{\langle Y_{nn}^{hs}\rangle}= T_{AB}^{hs} \sigma_{nn}^{inel}
\end{eqnarray}
This equation is similar to the original definition~Eq.~\ref{eq:3},
except that a different n-n overlap function is used.

A few words about $N_{coll}^{hs}$ are in order. First, according to
Eq.~\ref{eq:6}, there is no bias for minimum bias A-B event
selection, i.e. $\langle N_{coll}^{hs}\rangle=\langle
N_{coll}\rangle=\frac{AB\sigma_{nn}^{inel}}{\sigma_{AB}^{geo}}$.
Secondly, the n-n hard-parton overlap function $t_{hs}$ depends on
the $x$ and $Q^2$, thus in general, $N_{coll}^{hs}$ could depend on
the $p_T$. Last but not the least, the magnitude of the
hard-scattering cross-section $\sigma_{nn}^{hs}$ in
Eq.~\ref{eq:app2} is not important, what is the important is it's
spatial distribution. In fact, $\sigma_{nn}^{hs}$ could be
arbitrarily small, and yet would still cancel between A-B and n-n
(Eq.~\ref{eq:app5}). For our MC Glauber calculation, a convenient
choice is to choose $\sigma_{nn}^{hs}=\sigma_{nn}^{inel}$, then
Eq.~\ref{eq:app2} would directly give $N_{coll}^{hs}$.

\end{document}